\documentclass[english,prd,nofootinbib,preprintnumbers,twocolumn,showpacs]{revtex4-2}
\usepackage[latin1]{inputenc}
\usepackage{lmodern}
\usepackage{amsmath}
\usepackage{amsfonts}
\usepackage{appendix}
\usepackage{amssymb}
\usepackage{epsfig}
\usepackage{graphics,psfrag,rotating}
\usepackage{graphicx}
\usepackage{dcolumn}
\usepackage{bm}
\usepackage{epstopdf}
\usepackage{color}
\usepackage[usenames,dvipsnames,svgnames]{xcolor}
\usepackage[colorlinks=true,
            linkcolor=red,
            urlcolor=gray,
            citecolor=blue]{hyperref}
\usepackage{hyperref}
\usepackage[T1]{fontenc}
\usepackage{multirow}
\usepackage{float}
\usepackage{cancel}
\usepackage{hyperref}
\usepackage{adjustbox}
\usepackage{subfigure}

\usepackage{enumitem}

\def\3nab{\tilde{\nabla}}

\def\be {\begin{equation}}
\def\ee {\end{equation}}
\def\ba {\begin{align}}
\def\ea {\end{align}}

\def\bc {\begin{center}}
\def\ec {\end{center}}
\def\case#1/#2{\frac{#1}{#2}}

\newcommand{\bver}{\begin{verbatim}}
\newcommand{\ever}{\end{verbatim}}

\newcommand{\bea}{\begin{eqnarray}}
\newcommand{\eea}{\end{eqnarray}}
\newcommand{\beaa}{\begin{eqnarray*}}
\newcommand{\eeaa}{\end{eqnarray*}}

\newcommand{\Od}{{\mathcal O}}
\newcommand{\lsim}   {\mathrel{\mathop{\kern 0pt \rlap
  {\raise.2ex\hbox{$<$}}}
  \lower.9ex\hbox{\kern-.190em $\sim$}}}
\newcommand{\gsim}   {\mathrel{\mathop{\kern 0pt \rlap
  {\raise.2ex\hbox{$>$}}}
  \lower.9ex\hbox{\kern-.190em $\sim$}}}
\def\case#1/#2{\textstyle\frac{#1}{#2}}

\begin{document}

\title{TDiff invariant field theories for cosmology}

\author{Antonio L.\ Maroto}
\email{maroto@ucm.es}
\affiliation{Departamento de F\'{\i}sica Te\'orica and\\ Instituto de F\'{\i}sica de Part\'{\i}culas y del Cosmos (IPARCOS-UCM), Universidad Complutense de Madrid, 28040 
Madrid, Spain}




\begin{abstract} 
We study scalar field theories invariant under transverse diffeomorphisms 
in cosmological contexts. We show that in the geometric optics approximation, the corresponding particles move along geodesics and contribute with the same active mass (energy) to the gravitational field as in Diff invariant theories. However, for low-frequency (super-Hubble) modes, the contributions to the energy-momentum tensor differ from that of Diff invariant theories. This opens up a wide range of possibilities for cosmological model building.  As an example, we show that the simplest TDiff invariant scalar field theory with only kinetic term could drive inflation  and generate a nearly scale invariant (red-tilted) spectrum of density fluctuations. We also present a detailed analysis of cosmological perturbations and show that the breaking of full Diff invariance generically induces new non-adiabatic pressure perturbations.  A simple scalar field dark matter model based on a purely kinetic term  that exhibits the same clustering properties as standard cold dark matter is also presented. 
\end{abstract} 

\maketitle

\section{Introduction}

In recent years, interest in gravitational theories with broken diffeomorphisms invariance has grown. The most popular examples being the so-called unimodular gravity theories \cite{Einstein, Unruh:1988in, Alvarez:2005iy, Carballo-Rubio:2022ofy}. Unimodular  gravity restricts  the determinant of the metric tensor to be a non-dynamical field thus breaking the full diffeomorphisms (Diff) invariance of 
General Relativity down to transverse diffeomorphisms (TDiff) \cite{Alvarez:2006uu, Lopez-Villarejo:2010uib}.  The corresponding field equation are the trace-free Einstein equations in which vacuum, or in general any cosmological constant type contribution in the energy-momentum tensor, does not gravitate, thus providing a neat solution of the vacuum-energy problem \cite{Ellis:2010uc} (see however \cite{Padilla:2014yea,Jirousek:2023gzr} for a critical
point of view). In addition, if the energy-momentum tensor of the matter fields is conserved, it can be seen that
at the classical level,  unimodular gravity is equivalent to Einstein General Relativity supplemented with a cosmological constant term which appears as an integration constant. As a matter of fact, it has been shown that TDiff invariance is the minimal symmetry required by unitarity in theories with a massless spin 2 field \cite{vanderBij:1981ym}. Notice however that there are generally covariant formulations of unimodular gravity which in general introduce additional fields \cite{Henneaux:1989zc,Kuchar:1991xd,Jirousek:2018ago,Hammer:2020dqp}. More general gravity theories invariant under TDiffs but involving a dynamical metric determinant has been analyzed in \cite{Pirogov:2009hr,Pirogov:2011iq} and their potential implications for the cosmological dark sector have been studied in    \cite{Pirogov:2005im,Pirogov:2014lda}. Inflation models based on unimodular gravity have also  been considered in \cite{Ellis:2013uxa, Alvarez:2021cxy,Leon:2022kwn}.

Apart from the breaking of Diff invariance in the gravity sector,  theories with broken diffeomorphisms in the matter sector have also been analyzed \cite{Alvarez:2009ga,Jirousek:2020vhy,Dalmazi:2020xou}. Thus in \cite{Alvarez:2009ga} TDiff invariant models for spin 0 fields were explored. The phenomenology of such theories
included also the interesting possibility of degravitating the potential energy density \cite{Alvarez:2007nn}, but  pointed to  potential difficulties related to differences between inertial, active or passive masses and possible violations of the weak equivalence principle. In this work we revisit the viability of this type of  models and find the corresponding consistency conditions. 
In particular, we find that in the geometric optics approximation, viable TDiff models behave as Diff models from
a phenomenological point of view, however for very low frequencies, for instance, for super-Hubble modes in cosmological contexts, the behaviour of these models differs 
drastically from that of standard Diff theories. Thus, the scaling
of the energy density or the evolution of density perturbations
depend on the functions inducing the Diff invariance breaking. 
This different scaling properties opens a wide range  of possibilities for model building in cosmology. Thus, as  examples,  we present inflationary models driven by purely canonical kinetic terms which generate a red-tilted spectrum of field fluctuations or dark matter models without potential terms whose density perturbation behave as those of cold dark matter. 

The paper is organized as follows. In Section II we introduce the class of TDiff invariant scalar field theories that we will consider in the rest of the work. In Section III we discuss the covariant formalism for TDiff theories. 
 In Section IV, we study the model consistency. We perform the geometric optics approximation, develop its canonical quantization and calculate the active gravitational mass of single-particle states.
Sections V  and VI are devoted to  TDiff scalars in cosmological backgrounds. We calculate the corresponding energy-momentum tensor and  field equations and analyze the behaviour in three important regimes, namely, kinetic domination, potential domination and rapid oscillations. In Section VII, we consider the case of spaces with curvature and Section VIII is devoted to the construction of accelerated solutions and in particular to slow-roll inflationary models.
As an example, in Section IX, we compute the spectrum of primordial quantum fluctuations of a simple inflationary model
without potential term. In Section X, we perform the study of cosmological perturbations in TDiff models and pay attention to 
the evolution of adiabatic perturbations in models without potential term. In section XI, as an example, we show a simple model for dark matter. Finally in Section XII we present the main conclusions of the work. In an Appendix we introduce the basic concepts and definitions of TDiff transformations.

\section{TDiff invariant scalar field theories}

Let us consider the most general expression up to second order in 
derivatives for a minimally coupled scalar field action which is 
invariant under transverse diffeomorphisms 
\begin{align}
S_{\phi}=\int d^4x\, {\cal L}=  \int d^4x\left(\frac{f_K(g)}{2}g^{\mu \nu}\partial_\mu\phi\,\partial_\nu\phi-f_V(g)V(\phi)\right)\label{action}
\end{align}
with $f_K(g)$ and $f_V(g)$ arbitrary (positive) functions of the metric determinant. 
We will work with metric signature $(+,-,-,-)$. 
For $f_K(g)=f_V(g)=\sqrt{g}$ we recover the full diffeomorphisms invariance (see Appendix).

The equation of motion for the scalar field is
\begin{align}
\partial_\nu(f_K(g) g^{\mu \nu}\partial_\mu\phi)+f_V(g)V'(\phi)=0\label{KG}
\end{align}
whereas the energy-momentum tensor is defined in the usual way
as
\begin{align}
T^{\mu \nu}=-\frac{2}{\sqrt{g}}\frac{\partial {\cal L}}{\partial g_{\mu\nu}}
\end{align}
and reads\begin{align}
T_{\mu \nu}&=\frac{f_K(g)}{\sqrt{g}}\left(\partial_\mu\phi\, \partial_\nu\phi
-\frac{d\ln f_K}{d \ln g} g_{\mu\nu} g^{\alpha\beta}\partial_\alpha\phi\, \partial_\beta\phi\right)\nonumber \\
&+2\frac{f_V(g)}{\sqrt{g}}\frac{d\ln f_V}{d \ln g} g_{\mu\nu}V(\phi)
\end{align}
Notice that, because of the breaking of Diff invariance, in principle, $T^{\mu\nu}$ is not
 conserved on the solutions of the field equations of motion. Also notice that the Hilbert energy-momentum tensor that we have just obtained differs from the canonical one. 

Let us consider the total action for the scalar field coupled to gravity
\begin{align}
S&=S_{EH}+S_\phi=-\frac{1}{16\pi G}\int d^4x \sqrt{g}\,R\nonumber  \\
&+ \int d^4x\left(\frac{f_K(g)}{2}g^{\mu \nu}\partial_\mu\phi\,\partial_\nu\phi-f_V(g)V(\phi)\right) \label{actionTOT}
\end{align}
where only the scalar sector breaks Diff invariance down to TDiff. 
The corresponding  Einstein equations are obtained by considering 
(unconstrained) variations of the action with respect to the metric tensor
\begin{align}
R^{\mu\nu}-\frac{1}{2}g^{\mu\nu}R=8\pi G T^{\mu\nu}
\end{align}
Notice that Bianchi identities $G^{\mu\nu}_{\;\;\;\; ;  \nu}=0$ imply $T^{\mu\nu}_{\;\;\;\; ;  \nu}=0$, i.e. the energy-momentum is conserved  on the solutions of Einstein equations.

\section{Covariantized action}\label{SecII}
Although a general action
\begin{align}
S[g_{\mu\nu},\phi]=\int d^4x f(g){\cal L}(g_{\mu\nu},\phi,\partial_\mu\phi) \label{Taction}
\end{align}
 breaks Diff invariance down to TDiff, it is always possible to write it in
a generally covariant way \cite{Henneaux:1989zc,Kuchar:1991xd, Pirogov:2014lda} by introducing a new scalar density $\bar\mu$ which transforms 
as $\sqrt{g}$ under coordinate transformations. Thus,
\begin{align}
S[g_{\mu\nu},\phi,\bar \mu]=\int d^4x \sqrt{g}{\cal L}(g_{\mu\nu},\bar \mu/\sqrt{g},\phi,\partial_\mu\phi)\label{Diffaction}
\end{align}
can be seen to be equivalent to \eqref{Taction} in coordinates where $\bar \mu=1$. 
Thus for instance, following \cite{Henneaux:1989zc}, we can introduce the density $\bar \mu=\partial_\mu(\sqrt{g} \,T^\mu)$ where $T^\mu$ is a vector field under general diffeomorphisms. In this case, the action \eqref{actionTOT} can be written in the following covariant form
\begin{align}
S&=-\frac{1}{16\pi G}\int d^4x \sqrt{g}\,R \\
&+ \int d^4x \sqrt{g}\left(\frac{H_K(Y)}{2}g^{\mu \nu}\partial_\mu\phi\,\partial_\nu\phi-H_V(Y)V(\phi)\right)\nonumber 
\end{align}
where  we have introduced the scalar field
\begin{align}
Y=\nabla_\mu T^\mu
\end{align}
and 
\begin{align}
H_K(Y)&=Yf_K(Y^{-2})\\
H_V(Y)&=Yf_V(Y^{-2})
\end{align}
Notice that in this notation, the Diff case $f_K=f_V=\sqrt{g}$ corresponds
to $H_K=H_V=1$. 

Variations with respect to the scalar field $\phi$ yield
\begin{align}
\nabla_\nu(H_K(Y)g^{\mu\nu}\partial_\mu\phi)+H_V(Y)V'=0 \label{scalarcov}
\end{align}
where once again, prime denotes derivative with respect to its argument.  
On the other hand,  variations with respect to the vector field $T^\alpha$ lead to
\begin{align}
\partial_\alpha\left(H_K'(Y)X-H_V'(Y)V\right)=0\label{Yeq}
\end{align}
where we have defined
\begin{align}
X=\frac{1}{2}g^{\mu \nu}\partial_\mu\phi\,\partial_\nu \phi
\end{align}

Finally, variations with respect to the metric tensor allow us to obtain the corresponding Einstein equations, with the following expression for the energy-momentum tensor
\begin{align}
T_{\mu\nu}&=H_K(Y)\partial_\mu\phi\partial_\nu\phi-g_{\mu\nu}(H_K(Y)X-H_V(Y)V)\nonumber \\ 
&+ Y\left[H_K'(Y)X-H_V'(Y)V\right]g_{\mu\nu}
\end{align}
where we have used  equation \eqref{Yeq}. 

The covariantized action can be useful to develop the Hamiltonian
formalism and to localize the true dynamical degrees of freedom of the theory. However 
this is beyond the scope of current work and will be presented elsewhere. In this paper we are mainly interested in the cosmological aspects so that  we will not consider  this extra scalar density and will work with the action in the original form \eqref{actionTOT}. 

\section{Model consistency}
The general action in \eqref{action} is free from ghost and gradient instabilities provided
$f_K(g)>0$. In addition,  $V''(\phi_0)f_V\geq 0$ prevents also tachyon instabilities
around $\phi=\phi_0$ \cite{Rubakov:2014jja}. However, as shown in  \cite{Alvarez:2009ga},  the breaking of diffeomorphisms invariance could have important phenomenological implications as the three different types of masses, namely, inertial mass, active gravitational mass and passive gravitational mass, could in principle 
 differ one from the other in this kind of models. A difference between inertial and passive gravitational mass would lead to a violation of the Weak Equivalence Principle (WEP), whereas a difference between active and passive gravitational masses would violate Newton's third law.  On the other hand, the presence of the $f_K(g)$ and $f_V(g)$ which are space-time dependent functions in the Lagrangian of the theory could lead to violations of Local Position Invariance (LPI). WEP, Newton's third law and LPI are 
 extremely well tested experimentally. Let us then analyze the conditions that these functions should
 satisfy in order to have phenomenologically viable theories.

\subsection{Geometric optics approximation}

Possible violations of the WEP could appear in these models since it is not guaranteed that in the geometric optics approximation scalar field particles follow geodesics of the space-time geometry. Following \cite{Alvarez:2009ga}, let us then consider the case of massive scalar field with
\begin{align}
V(\phi)=\frac{1}{2}m^2 \phi^2
\end{align}
so that the equation motion reads
\begin{align}
\partial_\nu(f_K(g) g^{\mu \nu}\partial_\mu\phi)+f_Vm^2 \phi=0\label{KGm}
\end{align}
and write the scalar field in the geometric optics approximation as
\begin{align}
  \phi=\Re\left\{(A+B\epsilon+\dots)e^{i\frac{\theta}{\epsilon}}\right\}
  \end{align}  
with $\theta(x)$ a real function, $A(x)$ and $B(x)$ complex amplitudes and $\epsilon$ a 
small expansion parameter. Defining $k_\mu=\partial_\mu\theta$ and substituting in the equation of motion \eqref{KGm} we get to leading order $\Od(1/\epsilon^2)$
\begin{align}
A \left(g^{\mu\nu}k_\mu k_\nu-\frac{f_V}{f_K}m^2\right)=0 
\end{align}
which implies 
\begin{align}
g^{\mu\nu}k_\mu k_\nu-\frac{f_V}{f_K}m^2=0 \label{geom}
\end{align}
Thus taking the covariant derivative  and taking into account that 
$m$ is constant, we get
\begin{align}
2 k^\mu k_{\mu ; \nu}-m^2\left(\frac{f_V}{f_K}\right)_{,\nu}=0
 \end{align} 
 so that if $f_K=f_V$ then 
 \begin{align}
 k^\mu k_{\mu ; \nu}=k^\mu k_{\nu ; \mu}=0
 \end{align}
which is nothing but the geodesics equation for a particle with $g^{\mu\nu}k_\mu k_\nu=m^2$ according to \eqref{geom}.
Thus we see that the condition for the inertial mass to agree with the passive gravitational mass
and therefore to satisfy the WEP is that the two functions $f_K(g)$ and $f_V(g)$ are equal (or to differ in a constant which can be absorbed in the mass definition).

Regarding LPI, there are two possible ways to test it, namely, through gravitational redshift experiments or tests of constancy of non-gravitational fundamental constants \cite{Will}. In the case of gravitational redshift experiments,  \eqref{geom} shows that the mode 
frequency agrees with the standard Diff case for $f_K=f_V$ so that particle frequency depends
only on the gravitational potentials as required by LPI. Similarly from \eqref{geom} possible variations
of the particle mass $m$ could only be generated by a 
non-constant ratio $f_V/f_K$. 

To next to leading order $\Od(1/\epsilon)$ we get, for arbitrary $f_K$ and $f_V$
\begin{align}
2k^\mu\partial_\mu A +\frac{A}{f_K}\partial_\mu(f_K g^{\mu\nu}k_\nu)=0
\end{align}
which can be written as
\begin{align}
\frac{1}{f_K}\partial_\mu(A^2f_K g^{\mu\nu}k_\nu)=\frac{1}{f_K}\partial_\mu(\sqrt{g}A^2\frac{f_K}{\sqrt{g}} k^\mu)=0
\end{align}
In other words, $\nabla_\mu(A^2\frac{f_K}{\sqrt{g}}k^\mu)=0$, i.e.  $A^2\frac{f_K}{\sqrt{g}}k^\mu$ is a conserved vector \footnote{Notice that even though $V^\mu=A^2\frac{f_K}{\sqrt{g}}k^\mu$ is just a vector density,  we define its "covariant" divergence by $\nabla_\mu V^\mu=\frac{1}{\sqrt{g}}\partial_\mu(\sqrt{g}V^\mu)$}, so that Stokes theorem implies that
\begin{align}
\int_\Sigma d\Sigma_\mu \,A^2\frac{f_K}{\sqrt{g}}k^\mu
\end{align}
does not depend on the spatial hypersurface $\Sigma$. Here $d\Sigma_\mu=n_\mu d\Sigma$ where $n_\mu$ is a unit vector orthogonal to the
$\Sigma$ hypersurface and $d\Sigma=\sqrt{g_\Sigma} \,d^3x$ is the volume element on $\Sigma$ with $g^\Sigma_{\mu\nu}$ the induced metric on the hypersurface. 
Thus, we find for the amplitude
\begin{align}
A=C\left(\frac{g}{g_\Sigma}\right)^{1/4}\frac{1}{\sqrt{f_K n_\mu k^\mu}}
\end{align}
with $C$ a constant.

Thus for example,  for space-time geometries in which $ds^2=dt^2+g^\Sigma_{ij}dx^i dx^j$
 with $n_\mu=(1,0,0,0)$ we have 
$d\Sigma_\mu=\sqrt{g}(d^3x,0,0,0)$, and accordingly
\begin{align}
\partial_0\int_\Sigma d^3x \sqrt{g} A^2\frac{f_K}{\sqrt{g}}k^0=0
\end{align}
for any given 3-volume. Thus, we can interpret $ A^2\frac{f_K}{\sqrt{g}}k^0$ as the particle number density, so that the only difference with respect to the Diff invariant case is the 
appearance of the $f_K/\sqrt{g}$ factor in the number density of particles.

\subsection{TDiff invariant inner product}

Given two complex functions $\phi_1$ and $\phi_2$
which solve the equation of motion \eqref{KGm}, we can construct a TDiff invariant inner product in the following way. Let us define
\begin{align}
J^\mu=-i{g^{\mu\nu}}(\phi_1\partial_\nu\phi^*_2-\partial_\nu\phi_1 \phi^*_2)
\end{align}
Using the equation of motion it is straightforward to see that this current satisfies
\begin{align}
\frac{1}{f_K}\partial_\mu(f_K J^\mu)=0
\end{align}
for arbitrary $f_K$ and $f_V$ and accordingly we can define an inner product as
\begin{align}
(\phi_1,\phi_2)=-i \int_\Sigma d\Sigma^\mu \frac{f_K}{\sqrt{g}} \phi_1\overset{\leftrightarrow}{\partial_\mu}\phi_2^* \label{inner}
\end{align}
which again does not depend on the spatial hypersurface $\Sigma$ on which it is evaluated. 

\subsection{Canonical quantization}
As we have seen in the previous section, preserving the weak equivalence principle requires
the same volume element for the kinetic and potential terms in the action so that in the rest of the work we will limit ourselves to the case
\begin{align}
f_K(g)=f_V(g)=f(g)
\end{align}

Let us denote by $\phi_{\vec p}$ a complete set of mode functions which are solutions of the 
scalar equation of motion 
\begin{align}
\partial_\nu(f(g) g^{\mu \nu}\partial_\mu\phi)+f(g) m^2 \phi=0
\end{align}
which are  canonically normalized 
\begin{align}
(\phi_{\vec p},\phi_{\vec k})=(2\pi)^3\delta^{(3)}(\vec p-\vec k)
\end{align}
with respect to the inner product we have introduced in \eqref{inner}.
In order to define particle states associated to the $\phi$ field we can thus expand
a general solution of the equation of motion as
\begin{align}
 \phi(\eta,\vec x)=\int \frac{d^3k}{(2\pi)^3}(a_k\phi_k+a_k^\dagger\phi_k^*)
 \end{align} 
 with the canonical commutation relations for the creation and annihilation operators
 \begin{align}
 [a_p,a^\dagger_k]=(2\pi)^3\delta^{(3)}(\vec p- \vec k)
 \end{align}
and construct the corresponding Fock space from the vacuum state defined 
as usual by
\begin{align}
a_k\vert 0\rangle =0, \;\;\; \forall \vec k
\end{align}
In the geometric optics approximation, the modes can be written as
\begin{align}
\phi_{\vec k}=A_k e^{-i\theta_k}=\left(\frac{g}{g_\Sigma}\right)^{1/4}\frac{1}{\sqrt{2f n_\mu k^\mu}}e^{-i\theta_k}
\end{align}
where the normalization constant has been chosen so that the modes are canonically normalized, and at the leading order in the geometric optics approximation we
have
\begin{align}
\partial_\mu\phi_{\vec k}=-ik_\mu \phi_{\vec k}
\end{align}
with $k_\mu$ satisfying \eqref{geom}. 
Notice that the inner product for the mode solutions in the geometric optics approximation reads
\begin{align}
(\phi_{\vec p},\phi_{\vec k})=\int_\Sigma d^3x \sqrt{g_\Sigma}n^\mu(p_\mu+k_\mu)\frac{f}{\sqrt{g}} \phi_{\vec p}\,\phi_{\vec k}^*
\end{align}

Let us compute the expectation value of the energy on a spatial hypersurface
defined by the normalized timelike vector $n_\mu$ on a 
single particle state\footnote{For spacetime geometries in which $ds^2=dt^2+g^\Sigma_{ij}dx^i dx^j$
 with $n_\mu=(1,0,0,0)$ the normalization of the single particle state is given by 
$\langle  \vec k\vert \vec k\rangle=2k^0 V$ with $V$ the coordinate volume.} $\vert \vec k\rangle$ 
\begin{align}
 E_k&=\int_\Sigma d^3 x \sqrt{g_\Sigma} \,\frac{\langle \vec k\vert :T^\mu_{\;\;\nu}n_\mu n^\nu:\vert \vec k\rangle}{\langle  \vec k\vert \vec k\rangle}= \int_\Sigma d^3 x\,
f \sqrt{\frac{g_\Sigma}{g}}\nonumber\\
&\times\frac{\langle \vec k\vert :
n^\mu n^\nu \partial_\mu\phi\partial_\nu\phi -\frac{d\ln f}{d \ln g}\left(g^{\alpha\beta}\partial_\alpha\phi\, \partial_\beta\phi -m^2\phi^2\right):\vert \vec k\rangle}{\langle  \vec k\vert \vec k\rangle}\nonumber\\
&=n^\mu k_\mu \label{Eeikonal}
 \end{align} 
 which is the same expression we would have obtained in the Diff invariant case. 
 This is due to the fact that in the geometric optics approximation, the expectation value of the term multiplying
 $d\ln f/d\ln g$, which is nothing but the Lagrangian density, vanishes, i.e. $\langle  \vec k\vert :{\cal L}:\vert \vec k\rangle=0$ using \eqref{geom}.
  
In  the non-relativistic limit we can write $k_\mu=m \,n_\mu$. In this case,  we  obtain from \eqref{Eeikonal} that the 
 active gravitational mass contributing to the energy density, which is the source of the gravitational potential, would be simply $m$.

In conclusion for TDiff invariant field theories in the geometric optics approximation with $f_K(g)=f_V(g)$, inertial, active and passive gravitational mass are the same and agree with the standard Diff invariant expressions. This also guarantees that local position invariance is not violated.

\section{TDiff scalars  in cosmological backgrounds}
Let us consider the simple scenario of scalar fields in a Robertson-Walker background. Since  it is not possible
in general to fix coordinates in which $g_{00}=1$
with a TDiff transformation, we have to consider
a general form of the spatially homogeneous and isotropic Robertson-Walker metric \cite{Alvarez:2007nn}. We will work with flat spatial sections for simplicity
\begin{align}
ds^2=b^2(\tau)d\tau^2-a^2(\tau)d\vec x^2 \label{RW}
\end{align}
where now both $a(\tau)$ and $b(\tau)$ have to be obtained from  the  Einstein equations.

In this case, the equation of motion for the scalar field reads in the 
$f_K=f_V=f$ case
\begin{align}
 \phi '' +\frac{L'}{L} \phi '-\frac{b^2}{a^2}\vec\nabla^2\phi+b^2\frac{dV}{d\phi}=0
\end{align}
with $' =d/d\tau$ and 
\begin{align}
L(\tau)=\frac{f(g(\tau))}{b^2(\tau)}
\end{align}
where $g(\tau)=a^6(\tau) \,b^2(\tau)$. 
 
The energy-momentum tensor components read in this case
\begin{align}
T^0_{\;\;0}&=\frac{f(g)}{b^2\sqrt{g}}\left(
{\phi '}^2-\frac{d\ln f}{d \ln g}\left({\phi '}^2-\frac{b^2}{a^2}(\vec\nabla \phi)^2-2b^2V(\phi)\right)\right)\nonumber\\
T^i_{\;\;j}&=\frac{f(g)}{b^2\sqrt{g}} 
\left(-\frac{b^2}{a^2}\partial_i\phi\partial_j\phi \right.\nonumber \\
&\left.-\delta^i_{\; j}\frac{d\ln f}{d \ln g} \left({\phi '}^2-\frac{b^2}{a^2}(\vec\nabla \phi)^2-2b^2V(\phi)\right)\right)\nonumber  \\
T^0_{\;\;i}&=\frac{f(g)}{b^2\sqrt{g}}\phi'\partial_i\phi 
\end{align}

In order to study the contribution of TDiff scalar particles to the energy-momentum tensor, and in particular to obtain the active gravitational mass of non-relativistic
particles, we will limit ourselves to the case of a massive scalar field with $V(\phi)=\frac{1}{2}m^2\phi^2$. The corresponding equation of motion in Fourier space reads 
\begin{align}
 \phi''+\frac{L'}{L}\phi'+\frac{b^2}{a^2}k^2\phi+b^2m^2\phi=0
 \end{align} 
It is now  useful to introduce
\begin{align}
\phi=\frac{1}{\sqrt{L}}\hat\phi
\end{align}
with
\begin{align}
L=\frac{f}{b^2}
\end{align}
 so that 
the equation of motion reads
\begin{align}
\hat \phi''+\left(\frac{b^2}{a^2}(k^2+a^2m^2) +\frac{1}{4}\left(\frac{L'}{L}\right)^2-\frac{1}{2}\frac{L''}{L}\right)\hat \phi=0
\end{align}
Let us define
\begin{align}
\omega^2=\frac{b^2}{a^2}(k^2+a^2m^2)
\end{align}
In the high-frequency regime where the oscillation frequency is much larger that 
expansion rate, i.e. $\omega^2 \gg\{ (L'/L)^2,(L''/L)\}$  particle states can be properly 
defined. This  condition corresponds to the sub-Hubble regime and
is nothing but the geometric optics approximation for the Robertson-Walker background.
In this regime,  we can obtain the corresponding WKB mode solutions given
to leading adiabatic order by \cite{Birrell}
\begin{align}
\phi_k(\tau,\vec x)=\frac{1}{\sqrt{2L\omega}}e^{-i\int \omega d\tau+i\vec k\vec x} \label{WKB}
\end{align}
These modes are  canonically normalized with respect to the inner product we have defined in \eqref{inner}.
Following the results in the previous section, the expectation value of the energy on a
single particle state $\vert \vec k\rangle$ reads
\begin{align}
 E_k&=\int_\Sigma d^3 x \sqrt{g_\Sigma} \,\frac{\langle \vec k\vert :\rho:\vert \vec k\rangle}{\langle  \vec k\vert \vec k\rangle}= \int_\Sigma d^3 x\,
 \frac{a^3f}{b^2\sqrt{g}}  \\
&\times\frac{\langle \vec k\vert :\left(
{\phi '}^2-\frac{d\ln f}{d \ln g}\left({\phi '}^2-\frac{b^2}{a^2}(\vec\nabla \phi)^2-b^2 m^2\phi^2\right)\right):\vert \vec k\rangle}{\langle  \vec k\vert \vec k\rangle}\nonumber
 \end{align} 
where $\Sigma$ denotes the spatial hypersurface of constant $\tau$ time. 
Now substituting  the field expression in the WKB approximation \eqref{WKB}, we find to leading order
\begin{align}
E_k=\left(\frac{k^2}{a^2}+m^2\right)^{1/2}+\dots
\end{align}
Thus we see that for TDiff invariant theories in the geometric optics  approximation,  we recover the standard expression for the energy of  a particle in an expanding universe independently of the $f(g)$ function. In particular,  for ultrarrelativistic particles the energy of the particle scales as $1/a$ as in the Diff invariant case, so that $\rho_R\propto a^{-4}$.  On the 
other hand, for non-relativistic particles, we find that the contribution is just $m$ so 
that $\rho_M\propto a^{-3}$ and, as  mentioned before, the three types of masses (inertial, active and passive) agree with the Diff invariant case. The only condition being that the kinetic and potential functions are the same $f_K(g)=f_V(g)$ and this condition also guarantees that TDiff scalar
particles also propagate along geodesics.  However, as we will see in next section, for homogeneous fields or in general for super-Hubble modes, i.e. far from the geometric optics approximation, the energy density of TDiff fields deviate from the standard Diff behaviour.

\section{Homogeneous scalar fields}

Let us then consider the simplest scenario with homogeneous scalar fields $\phi=\phi(\tau)$ in a Robertson-Walker background.  
In this case, the equation of motion for the scalar field reads
\begin{align}
 \phi '' +\frac{L'}{L} \phi '+b^2\frac{dV}{d\phi}=0 \label{phi}
\end{align}
Notice that the solution of the field equations for $V(\phi)=0$ is simply 
\begin{align}
\phi '(\tau)=\frac{C_\phi}{L(\tau)}
\end{align}
with $C_\phi$ a constant.

The energy-momentum tensor reads
\begin{align}
\rho=T^0_{\;\;0}&=\frac{f(g)}{b^2\sqrt{g}}\left(1-\frac{d\ln f}{d \ln g}\right){\phi '}^2+2\frac{f(g)}{\sqrt{g}}\frac{d\ln f}{d \ln g}V(\phi)\label{rho}\\
p=-\frac{1}{3} T^i_{\;\;i}&=\frac{f(g)}{b^2\sqrt{g}} \frac{d\ln f}{d \ln g} {\phi '}^2-2\frac{f(g)}{\sqrt{g}}\frac{d\ln f}{d \ln g}V(\phi)\label{p}
\end{align}
Notice that in the case in which $f(g)=$ const. the potential terms
do not contribute to the energy-momentum tensor and  the pressure is identically zero so that 
the scalar field behaves as non-relativistic matter. In the opposite case
with $\vert\frac{d\ln f}{d \ln g}\vert\gg 1$, we get  $p=-\rho$ and 
the field behaves as a cosmological constant, regardless the concrete field evolution. 

The corresponding  Einstein equations read
\begin{align}
\frac{a'^2}{a^2}&= \frac{8\pi G}{3}\rho\, b^2\\
2\frac{a''}{a}+ \frac{a'^2}{a^2}-2\frac{a'  b'}{ab}&=-8\pi G p\, b^2
\end{align}

The conservation equation $T^{\mu\nu}_{\;\;\;\; ;  \nu}=0$ implies
\begin{align}
\rho ' +3\frac{a'}{a}(\rho +p)=0
\end{align}

In cosmological time $dt=b(\tau)d\tau$, Einstein equations read
\begin{align}
 \frac{\dot a^2}{a^2}&= \frac{8\pi G}{3}\rho \label{Friedmann}\\  
\frac{2 \ddot a}{a}+ \frac{\dot a^2}{a^2}&=-8\pi G\, p 
\end{align}
with $\dot{}=d/dt$, i.e. we recover the standard form where the Hubble parameter corresponds to $H=\dot a/a$  and the conservation equation also takes the usual form
\begin{align}
\dot\rho  +3H(\rho +p)=0 \label{cons}
\end{align}

On the other hand, the field equation can be recast as
\begin{align}
\ddot \phi+\frac{\dot J}{J}\dot \phi+V'(\phi)=0\label{KGRW}
\end{align}
where prime denotes derivative with respect to the argument and 
\begin{align}
J=Lb=\frac{f}{b}=a^3\frac{f}{\sqrt g}
\end{align}

A few comments are in order at this point. Even though  Einstein and conservation equations  take the usual form in cosmological time, it is not possible to set $b=1$ from the very beginning. This is so  because if we had set $b=1$ in the the energy-momentum tensor components  \eqref{rho} and \eqref{p}, it would not have been possible to 
solve simultaneously the field equation \eqref{KGRW} and the conservation equation \eqref{cons}.
The compatibility of these two equations imposes a condition on $b(\tau)$.  In general it is not possible to obtain explicit expressions for $b(\tau)$, however there are three limiting cases in which 
the consistency condition can be worked out.

\subsubsection{Potential domination}
When the kinetic term is negligible, $p+\rho=0$ and therefore the field equation of state is
\begin{align}
w_\phi=\frac{p}{\rho}= -1
\end{align}
as in the Diff invariant case. The conservation equation \eqref{cons} 
thus requires $\rho=$ const. Since the field equation of motion in this case implies also $\phi=$ const.
the conservation equation translates into 
\begin{align}
\frac{f(g)}{\sqrt{g}}\frac{d\ln f(g)}{d\ln g}=C_g
\end{align}
with $C_g$ an arbitrary constant. Thus we see that for the particular  models with $f(g)=C_0+2C_g\sqrt{g}$ 
with $C_0$ an arbitrary constant, energy-momentum tensor is automatically conserved
on solutions of the equations of motion  so that $a$ and $b$ are independent functions.  On the other hand, for arbitrary $f(g)$, the above equation implies
\begin{align}
g=\text{const.}
\end{align}
i.e.  $b\propto a^{-3}$.

\subsubsection{Kinetic domination}
For $V(\phi)=0$  the equation of state is given by
\begin{align}
 w_\phi=\frac{\frac{d\ln f}{d \ln g}}{1-\frac{d\ln f}{d \ln g}} \label{eqstate}
 \end{align} 
Thus, for simple power law models with $f(g)\propto g^\alpha$ we have 
 \begin{align}
 w_\phi=\frac{\alpha}{1-\alpha}\label{wa}
 \end{align}
 which recovers the standard stiff fluid behaviour with $w_\phi=1$ for the Diff invariant case $\alpha=1/2$. 
 
 For exponential functions
 $f\propto e^{-\beta g}$ with $\beta>0$ constant, the equation of state reads
  \begin{align}
  w_\phi=-\frac{\beta g}{1+\beta g}
  \label{exp}
 \end{align}
which interpolates between $ w_\phi=0$ for $\beta g \ll 1$ and 
$w_\phi=-1$ for   $\beta g \gg 1$. On the other hand for $f\propto e^{\beta/g}$, the equation of state reads
 \begin{align}
 w_\phi=-\frac{1}{1+\frac{g}{\beta}}
 \end{align}
which interpolates between $w_\phi=-1$ for $ g/ \beta \ll 1$ and 
$w_\phi=0$ for   $ g/\beta \gg 1$.

Using the expressions for $\rho$ and $p$ and the equation  of motion \eqref{phi}, the conservation equation can be integrated so that
\begin{align}
\frac{g}{f(g)}\left(1-2\, \frac{d\ln f(g)}{d\ln g}\right)=C_g \,a^{6} \label{ab}
\end{align}
with $C_g$ an integration constant. This equation allows to obtain $b(\tau)$ as 
a function of $a(\tau)$ for a given $f(g)$. The integration constant $C_g$
can be fixed by imposing $a=b=1$ today. Notice that for $f(g)=\sqrt{g}$ the left hand side of \eqref{ab} vanishes so that $b$ and $a$ are independent functions, as expected for a Diff invariant theory\footnote{A derivation of this expression from a  covariantized action can be found in the Appendix} .

For example, for power-law models $f\propto g^\alpha$ this condition implies 
\begin{align}
 g\propto a^{\frac{6}{1-\alpha}} \label{gaKin}
 \end{align} 
 which translates into $b\propto a^{\frac{3\alpha}{1-\alpha}}$. 
 On the other hand, this implies that $L=f(g)/b^2=$ const. and accordingly   
\begin{align}
\phi'=\text{const}. \label{phiprima}
 \end{align} 
For the $J$ function we obtain in this case 
\begin{align}
J\propto b \propto a^{\frac{3\alpha}{1-\alpha}} \label{J}
\end{align}
In addition, the scaling of the determinant \eqref{gaKin} means that for $w_\phi>-1$ ($\alpha<1$) $g$  grows with the scale factor, whereas
it is decreasing for $w_\phi<-1$ ($\alpha>1$). Notice however that positive energy density in this case requires $\alpha < 1 $. Thus in the exponential case discussed above \eqref{exp} 
in which $w_\phi>-1$,  $\beta g$ is a growing function of $a$, i.e. we expect a natural evolution in time from $w_\phi=0$ to $w_\phi=-1$. In particular, in order to have today $w_\phi=-\beta/(1+\beta)=-\Omega_\Lambda$
as in $\Lambda$CDM, we would require $\beta=\Omega_\Lambda/\Omega_M$.

 As a consistency check we also derive the  condition \eqref{ab} from the covariantized equations
 in Section \ref{SecII} just by imposing that in coordinates in which $\bar \mu=1$, we have 
\begin{align}
Y=\frac{1}{\sqrt{g}}=\frac{1}{a^3b} \label{Ymu1}
\end{align}
 Thus, in this case, \eqref{scalarcov}, reads
\begin{align}
\frac{d}{d \tau}\left(H_K \frac{a^3}{b}\frac{d\phi}{d\tau}\right)=0
\end{align}
whereas \eqref{Yeq} implies
\begin{align}
\frac{d}{d \tau}\left( \frac{H_K'}{b^2}\left(\frac{d\phi}{d\tau}\right)^2\right)=0
\end{align}
Thus, combining both equations we get
\begin{align}
H_K^2(Y)a^6= C H_K'(Y)
\end{align}
with $C$ a constant.  Using \eqref{Ymu1} we recover \eqref{ab} in a
straightforward way.

\subsubsection{Rapidly oscillating evolution}
So far we have considered either potential or kinetically dominated evolution, now we will consider 
the phase of rapid oscillating around the potential minimum which is typically present in reheating
scenarios after inflation or in ultralight dark matter models.  In this case, both potential and kinetic terms contribute to the energy-momentum tensor but if the oscillations are fast compared to the expansion rate,  the conditions on the metric tensor in order to satisfy the average energy-momentum tensor conservation equation can also be obtained explicitly.

From the equation of motion \eqref{KGRW} and following \cite{Turner:1983he,Johnson:2008se,Cembranos:2015oya}, we can see that if the frequency of field oscillations $\omega$ is large compared to the universe expansion rate, i.e
$\omega\gg \dot J/J$, and provided
the amplitude of the field 
oscillations is bounded, then the average over a time  interval $T\gg \omega^{-1}$ satisfies
\begin{align}
\left\langle \frac{d}{dt}(\dot \phi \phi)\right\rangle \ll \langle \dot \phi^2\rangle  \label{neg}
\end{align}
Thus, using the equation of motion, neglecting $\dot J/J$ terms, we can write
\begin{align}
\langle \dot \phi^2+ \ddot \phi \phi\rangle= \langle \dot \phi^2 -V' \phi\rangle
\end{align}
so that using \eqref{neg}, we get
\begin{align}
\langle \dot \phi^2\rangle= \langle V' \phi\rangle \label{avkin}
\end{align}
On the other hand, for $f\propto g^\alpha$ we can write for the average equation of state
\begin{align}
 \langle w_\phi\rangle =\frac{ \langle p\rangle }{ \langle \rho\rangle }=\frac{\alpha \langle\dot \phi^2\rangle-2 \alpha\langle V\rangle}{(1-\alpha)\langle\dot\phi^2\rangle+2\alpha \langle V\rangle}
 \end{align} 
 For power-law potentials $V(\phi)\propto \phi^n$,  using \eqref{avkin} we have $\langle \dot \phi^2\rangle=n\langle V\rangle$ and substituting in the previous expression we get
 \begin{align}
 \langle w_\phi\rangle =\frac{\alpha(n-2)}{n-\alpha(n-2)} \label{avw}
\end{align}  
 So that for the Diff invariant case $\alpha=1/2$ we recover the well-know 
 expression \cite{Turner:1983he} $ \langle w_\phi\rangle=(n-2)/(n+2)$. 

We see that unlike the Diff invariant case, in which the matter behaviour $\langle w_\phi\rangle=0$
is only realized for $n=2$, in  the TDiff case, this also happens for oscillations in any type of potential for $\alpha=0$. We also see that for $\vert\alpha \vert \gg 1$, 
$\langle w_\phi\rangle\simeq -1$  as expected. 
The radiation case  $\langle w_\phi\rangle=1/3$ which corresponds to $n=4$ in the Diff invariant case, now can be obtained for $n=8\alpha/(4\alpha-1)$.
 
In the case in which the potential is a simple mass term $V(\phi)=\frac{1}{2}m^2\phi^2$ it is 
possible to obtain the condition for the energy-momentum tensor conservation in an explicit way. 
In this case, the equation of motion \eqref{KGRW}  reads
\begin{align}
\ddot \phi+\frac{\dot J}{J}\dot \phi+m^2 \phi=0
\end{align}
Defining
\begin{align}
\phi(t)=\frac{1}{\sqrt{J}}\chi(t)
\end{align}
the equation of motion reads
\begin{align}
\ddot\chi+\omega^2(t)\chi=0
\end{align}
with
\begin{align}
\omega^2(t)=m^2+\frac{\dot J^2}{4J^2}-\frac{\ddot J}{2J}
\end{align}
For slowly varying frequency $\omega(t)$, i.e. $m^2\gg\{\ddot J/J,\dot J^2/J^2\}$, we can use the WKB approximation \cite{Birrell} to find solutions. Up to second adiabatic order we can write
\begin{align}
\chi(t)=\frac{C}{\sqrt{W(t)}}\cos\left(\int^t W(t')dt'\right)
 \end{align} 
with $C$ constant and 
\begin{align}
W^2(t)=\omega^2(t)-\frac{1}{2}\left(\frac{\ddot\omega}{\omega}-\frac{3}{2}\frac{\dot\omega^2}{\omega^2}\right)
\end{align}
Substituting in  the conservation equation \eqref{cons} and averaging
\begin{align}
\langle \dot\rho\rangle  +3H\langle \rho +p\rangle=0
\end{align}
we find that the equation is automatically satisfied to first order in the adiabatic approximation,
whereas to second order we find
 \begin{align}
&\frac{(2 \alpha -1)}{8m a^3} \left[
6\alpha\left(\frac{\ddot a}{a} \frac{\dot b}{b} -  \frac{\dot a^2}{a^2} \frac{\dot b}{b} + \frac{\dddot a}{a}+2 \frac{\dot a^3}{a^3}-3\frac{\dot a}{a} \frac{\ddot a}{a}\right) \right. \nonumber \\
&\left.+(2\alpha-1)\left(\frac{\dddot b}{b}+\frac{\dot b^3}{b^3}-2\frac{\dot b}{b}\frac{\ddot b}{b}\right)\right]=0
 \end{align} 
Again we see that the conservation equation is satisfied in the Diff case, whereas for $\alpha\neq 1/2$
we can find $b(t)$ as a function of $a(t)$. For power-law solutions $b(t)\propto a^p(t)$ we find
\begin{align}
p=\frac{6\alpha}{1-2\alpha}
 \end{align} 
so that 
\begin{align}
g=a^6 b^2 \propto a^{\frac{6}{1-2\alpha}}
\end{align}
 In turn, this implies that
$J=f/b\propto g^\alpha/b=$ const. so that in the TDiff case, the field $\phi$ oscillates with constant amplitude for any value of $\alpha$. This contrasts with the Diff invariant case in which 
the field oscillates with an amplitude decreasing as $a^{-1}$. Notice that even though the amplitude of
oscillations is constant, the energy density scales as $\rho\propto f/\sqrt{g}\propto a^{-3}$ as
expected for an average equation of state \eqref{avw} $ \langle w_\phi\rangle=0$ corresponding
to the mass case $n=2$.

\section{Spaces with curvature}
So far we have limited ourselves to Robertson-Walker metrics with 
vanishing curvature.  In order to analyze spaces with spatial curvature, we note that the 
metric determinant appearing in the definition of the energy-momentum tensor
could break homogeneity in certain coordinate systems thus preventing the 
existence of solutions in those cases. Thus for example, in spherical coordinates
\begin{align}
ds^2=b^2(\tau)d\tau^2-a^2(\tau)\left(\frac{dr^2}{1-Kr^2}+r^2d\theta^2+r^2\sin^2\theta d\phi^2\right)
\end{align}
we have $g=b^2a^6r^4\sin^4\theta/(1-Kr^2)$. In order to avoid this problem, 
following \cite{Coley:2006xu} we
introduce the change of coordinates
\begin{align}
u&=\cos\theta\\
d\hat r&=\frac{dr \,r^2}{\sqrt{1-Kr^2}}
\end{align}
so  that the metric reads
\begin{align}
ds^2&=b^2(\tau)d\tau^2 \nonumber \\
&-a^2(\tau)\left(\frac{d\hat r^2}{A^4(\hat r)}+A^2(\hat r)\frac{du^2}{1-u^2}+A^2(\hat r)(1-u^2) d\phi^2\right)
\end{align}
where
\begin{align}
\frac{dA(\hat r)}{d\hat r}=\frac{\sqrt{1-KA^2(\hat r)}}{A^2(\hat r)}
\end{align}
In these coordinates $g=b^2a^6$ is only function of time for any value of $K$. 

The corresponding  Einstein equations in these coordinates read
\begin{align}
\frac{a'^2}{a^2}+\frac{Kb^2}{a^2}&= \frac{8\pi G}{3}\rho\, b^2\\
2\frac{a''}{a}+ \frac{a'^2}{a^2}-2\frac{a'  b'}{ab}+\frac{Kb^2}{a^2}&=-8\pi G p\, b^2
\end{align}
so that in cosmological time we recover again the standard Friedmann equations
with curvature
\begin{align}
 \frac{\dot a^2}{a^2}+\frac{K}{a^2}&= \frac{8\pi G}{3}\rho \\
\frac{2 \ddot a}{a}+ \frac{\dot a^2}{a^2}+\frac{K}{a^2}&=-8\pi G\, p
\end{align}
where $\rho$ and $p$ are given by \eqref{rho} and \eqref{p}. As expected, the equation of motion for the scalar field takes the same form as in the $K=0$ case \eqref{KGRW}.

\section{Accelerated solutions}
In this section we will obtain accelerated expansion solutions of scalar \eqref{KGRW}
and Friedmann \eqref{Friedmann} equations in cosmological time which could 
be relevant either for early universe inflation or late-time acceleration. Unlike Diff invariant
models, as mentioned before, now it is possible to have accelerated expansion driven 
both by kinetic or potential terms. Let us then start with the simplest case of purely kinetically driven
acceleration 

\subsubsection{Kinetically driven acceleration}

In the $V(\phi)=0$ case, it is possible to have inflationary solutions
provided $w_\phi=p/\rho<-1/3$ which implies
\begin{align}
\frac{d\ln f}{d \ln g}<-\frac{1}{2} \label{inf}
\end{align}
In order to have inflationary solutions, we require functions $f(g)$ such that the 
condition \eqref{inf} is satisfied during a given period of time in the early universe
and it is then violated so that inflation can come to an end. Thus, a simple example would be 
\begin{align}
f(g)=\sqrt{g}+ \left(\frac{g}{g_0}\right)^\alpha  \label{power}
\end{align}
where $g_0$ is a positive constant. Imposing the condition \eqref{inf} we find that $\alpha<-1/2$ in order to get accelerated expansion. In this case, \eqref{ab} implies
\begin{align}
g\propto a^{\frac{6}{1-\alpha}}
\end{align}
i.e. $b\propto a^{\frac{3\alpha}{1-\alpha}}$ during the acceleration phase. Thus we see that the metric determinant is a monotonic growing function of $a$, so that for $\alpha<-1/2$ the second term
in \eqref{power} will become subdominant at late times, recovering the standard behaviour for the volume form and eventually ending inflation. The scale factor at the end of inflation is given by
\begin{align}
a_{end}\simeq \left(-\frac{1+2\alpha}{2g_0^\alpha}\right)^{\frac{1-\alpha}{3(1-2\alpha)}}
\end{align}
The equation of state is thus in the range
\begin{align}
\frac{\alpha}{1-\alpha}<w_\phi<1
\end{align}
The energy density during inflation when the second term in \eqref{power} dominates is given by
\begin{align}
 \rho\simeq C_\phi^2\,g_0^\alpha\, (1-\alpha)\,a^{-\frac{3}{1-\alpha}}
 \end{align}

\subsubsection{Slow-roll inflation}
Let us now consider the general case with a potential term. We will work again for simplicity in the power-law case $f\propto g^\alpha$. From Einstein equations we get
\begin{align}
\dot H=-4\pi G(\rho +p)=-4\pi G \frac{f}{\sqrt{g}}\dot\phi^2
\end{align}
and the corresponding slow-roll parameters \cite{Lyth,Riotto:2002yw} are
\begin{align}
 \varepsilon=-\frac{\dot H}{H^2}=4\pi G \frac{f}{\sqrt{g}}\frac{\dot\phi^2}{H^2}=\frac{3}{4\alpha}\frac{\dot\phi^2}{V} \label{epsilon}
 \end{align} 
 Thus the equation of state reads
 \begin{align}
 w_\phi=\frac{\alpha\dot \phi^2-2\alpha V}{(1-\alpha)\dot \phi^2+2\alpha V} 
 \label{omegaphi}
 \end{align}
 so that in the slow-roll approximation
 \begin{align}
 1+w_\phi=\frac{\dot\phi^2}{2\alpha V}=\frac{2}{3}\varepsilon
 \end{align}
 which is the same expression as in standard Diff invariant quintessence models.
 
 On the other hand, in the slow-roll approximation the equation of motion \eqref{KGRW} reads
 \begin{align}
 6\alpha\dot\phi H+(2\alpha-1)\frac{\dot b}{b}\dot \phi+V'\simeq 0 \label{kgslow}
 \end{align}
 Imposing the conservation of the energy momentum-tensor we find
 \begin{align}
 \dot \rho+3 H\frac{f}{\sqrt{g}}\dot\phi^2=0
 \end{align}
 so that in the slow-roll approximation we get
 \begin{align}
\left(\alpha-\frac{1}{2}\right) \frac{\dot g}{g}=\left(\left(6\alpha-\frac{3}{2\alpha}\right)H+(2\alpha-1)\frac{\dot b}{b}\right)\frac{\dot \phi^2}{V}
 \end{align}
which, as expected, is automatically satisfied  for  $\alpha = 1/2$, whereas for  $\alpha\neq 1/2$ we get to first order in the slow-roll parameters
\begin{align}
 \frac{\dot b}{b}=(2\varepsilon-3)H
 \end{align} 
 which implies $g\propto a^{4\varepsilon}$. Notice that this scaling implies that the prefactor $f/\sqrt{g}\propto a^{(4\alpha-2)\varepsilon}$ is also slowly evolving. 
 Substituting back into \eqref{kgslow}, we get
 \begin{align}
(3+(2\alpha-1)2\varepsilon)\dot\phi H+V'\simeq 0 
 \end{align}
  so finally we can write from \eqref{epsilon}
\begin{align}
 \varepsilon=\frac{1}{64\pi G\alpha^2}\frac{\sqrt{g}}{f}\left(\frac{V'}{V}\right)^2
 \end{align}  
 On the other hand, the second slow-roll parameter reads
 \begin{align}
 \delta=-\frac{\ddot \phi}{\dot\phi H}=\frac{V''}{H^2}-\varepsilon=-\varepsilon+\frac{V''}{16\pi G\alpha V}\frac{\sqrt{g}}{f}
 \end{align}
 so that 
 \begin{align}
 \eta=\varepsilon+\delta=\frac{V''}{16\pi G\alpha V}\frac{\sqrt{g}}{f}
 \end{align}
Notice that the appearance of the $\frac{\sqrt{g}}{f}$ extra factor in the slow-roll parameters opens the possibility for 
models with unsuitable potentials to support inflation  whenever $f(g)\gg\sqrt{g}$ at early times. 
 
 The corresponding number of e-folds is then given to leading order in the 
 slow-roll approximation  by 
\begin{align}
 N=\int_{t_i}^{t_f} \frac{\dot a}{a}dt=\int_{\phi_i}^{\phi_f}H\frac{d\phi}{\dot \phi}\simeq 
 -3\int_{\phi_i}^{\phi_f}\frac{H^2}{V'}d\phi \nonumber 
 \\=-16\alpha\pi G  \frac{f}{\sqrt{g}}\int_{\phi_i}^{\phi_f}\frac{V}{V'}d\phi
 \end{align} 
 where in the last step we have used that $f/\sqrt{g}$ is constant to first
 order in the slow-roll approximation. Notice also that again a large $f/\sqrt{g}$ factor will enhance the total number of e-folds generated. 
 
\section{Quantum fluctuations during inflation}
As an example of the previous results, we will compute the spectrum of quantum fluctuations 
for a massless scalar field driving inflation. With that purpose, let us consider the simplest TDiff inflation model
\begin{align}
S_{\phi}= \int d^4x f(g)\frac{1}{2}g^{\mu \nu}\partial_\mu\phi\,\partial_\nu\phi
\end{align}
with $f\propto g^\alpha$.  Let us consider scalar quantum fluctuations $\delta \phi$ around a classical homogeneous background field $\phi_0(t)$ so that
\begin{align}
\phi(t,\vec x)=\phi_0(t)+\delta\phi(t,\vec x)
\end{align}
Conservation of the energy-momentum tensor for the classical background implies \eqref{J}
 $J\propto  a^{\frac{3\alpha}{1-\alpha}}$. Accordingly we can write the equation for $\delta \phi$ 
in Fourier space as
\begin{align}
 \delta\ddot\phi_k+\frac{3\alpha}{1-\alpha}H \delta\dot\phi_k +\frac{k^2}{a^2}\delta\phi_k=0
\end{align}
We introduce the conformal time $dt=a(\eta)d\eta$ and redefine the field as 
\begin{align}
\delta\phi_k=\sqrt{\frac{a}{J}}\delta\hat \phi_k
\end{align}
The scale factor during inflation can be written as
\begin{align}
a(\eta)=(-C_I\eta)^{\frac{2}{1+3w_\phi}}
\end{align}
with $w_\phi=\alpha/(1-\alpha)$ and $C_I$ a constant  so that the Hubble parameter
can be written as
\begin{align}
H^2=\frac{4C_I^2}{(1+3w_\phi)^2}a^{-3-3w_\phi}
\end{align}
and the equation for the fluctuations reads
\begin{align}
\delta\hat\phi''_k+\left(k^2-\frac{1}{\eta^2}\left(\nu^2-\frac{1}{4}\right)\right)\delta\hat\phi_k =0
\end{align}
where prime denotes derivative with respect to the conformal time and 
\begin{align}
\nu=\left\vert\frac{3+3w_\phi-3\alpha(3+w_\phi)}{2(\alpha-1)(1+3w_\phi)}\right\vert=\frac{-3 + 6 \alpha}{2 + 4 \alpha}
\end{align}
where in the last step we have considered that $\alpha<-1/2$ in order to have accelerated 
expansion.

In order to define the Bunch-Davies vacuum we consider mode solutions that in the sub-Hubble regime ($k\gg aH$) behave as positive frequency modes with the correct normalization \eqref{WKB}, i.e.
\begin{align}
\delta\phi_k=\sqrt{\frac{a}{2Jk}} e^{-i k\eta}, \;\;\; \vert k\eta\vert \gg 1
\end{align}
with $k=\vert \vec k\vert$, which corresponds 
 to
\begin{align}
\delta\phi_k=\frac{\sqrt{a\pi}}{2\sqrt{Jk}}e^{i\left(\nu+\frac{1}{2}\right)\frac{\pi}{2}}(-k\eta)^{1/2}H_\nu^{(1)} (-k\eta)
 \end{align} 
 where $H_\nu^{(1)}$ are the Hankel function of first kind. In the super-Hubble regime  ($k\ll aH$) 
 these solutions behave as
 \begin{align}
\vert\delta\phi_k\vert=\sqrt{\frac{a}{J}}\frac{2^{\nu-3/2}}{\sqrt{2k}}\frac{\Gamma(\nu)}{\Gamma(3/2)}(-k\eta)^{\frac{1}{2}-\nu}, \;\;\; \vert k\eta\vert \ll 1 \label{superH}
 \end{align} 
 The corresponding spectrum is given by
 \begin{align}
 P_{\delta\phi}(k)=\frac{k^3}{2\pi^2}\vert \delta\phi_k\vert^2
 \end{align}
which can be written as 
\begin{align}
 P_{\delta\phi}(k)=C^2\left(\frac{H}{2\pi}\right)^2 a^{\frac{3-6\alpha}{1-\alpha}}(-k\eta)^{3-2\nu}\label{spectrum}
\end{align} 
with
\begin{align}
C=\frac{2^{\nu-3/2}\Gamma(\nu)}{\Gamma(3/2)}\frac{1+3w_\phi}{2}
\end{align}
and the spectral index $n_s$ is given by
\begin{align}
n_s-1=3-2\nu=\frac{6}{1 + 2 \alpha}
\end{align}
Thus we see that for TDiff massless inflaton, the acceleration condition $\alpha<-1/2$, implies that $n_S-1<0$, i.e. the spectrum of fluctuation is red-tilted. A nearly scale-invariant spectrum thus 
requires $\alpha=\frac{d\ln f}{d\ln g}\ll -1$.

Consider now the fluctuation in the field energy density contrast, given by
\begin{align}
\frac{\delta\rho_k}{\rho_0}=2\frac{\delta\dot\phi_k}{\dot\phi_0}=\frac{2J\delta\phi'_k}{Ca}
\end{align}
where we have used $\dot \phi_0=C/J$ with $C$ constant. Substituting the super-Hubble solution from \eqref{superH} we find that 
\begin{align}
\frac{\delta\rho_k}{\rho_0}=\text{const.}
\end{align}
i.e. even though the spectrum of field perturbations \eqref{spectrum} depends on time, 
we see that the spectrum of the density contrast is constant for super-Hubble modes, very much 
as in the Diff invariant case.

\section{Metric perturbations}
Let us consider the most general form of the flat Robertson-Walker metric
with scalar perturbations
 \begin{align}
ds^2&=b^2(\tau)(1+2\Phi)d\tau^2-2a^2(\tau)B_{,i}dx^id\tau \nonumber \\ 
&-a^2(\tau)[(1-2\Psi)\delta_{ij}+2E_{,ij}]dx^i dx^j
 \end{align}
 Under arbitrary infinitesimal diffeomorphisms $\hat x^\mu=x^\mu+\xi^\mu(x)$, the scalar 
 potentials transform as
 \begin{align}
 \hat \Phi&=\Phi-\frac{b'}{b}\xi^0-{\xi^0}'\\
\hat \Psi&=\Psi+\frac{a'}{a}\xi^0\\
\hat B&=B+\frac{b^2}{a^2}\xi^0-{\xi}' \label{B}\\
\hat E&=E-\xi \label{E}
 \end{align}
with $\xi^i=\partial_i \xi$. TDiff restricts the transformations to those which preserve
the metric determinant, i.e. $\hat g(\hat x)=g(x)$ which implies $\partial_\mu \xi^\mu=0$ or 
\begin{align}
{\xi^0}'+\nabla^2\xi=0 \label{cond}
\end{align}
Notice that this condition is different from that imposed for unimodular gravity in \cite{Gao:2014nia}. 
This means that, unlike the Diff invariant case, for the TDiff invariant theories we only have one
gauge freedom, so that the usual longitudinal gauge conditions $E=0$ and $B=0$ cannot be 
 simultaneously imposed. However it is always possible to set one  combination of the four scalar potentials to zero. In particular, as we will see below, it is useful to work in the $E'-B=0$ gauge. 

The condition \eqref{cond} means that the metric determinant 
\begin{align}
g=b^2a^6(1+2(\nabla^2 E+\Phi-3\Psi))
\end{align}
transforms as 
\begin{align}
\hat g(x)=g(x)-g(x)\left(\frac{b'}{b}+3\frac{a'}{a}\right)\xi^0
 \end{align} 
under TDiffs.

For the perturbed scalar field $\phi=\phi_0(\tau)+\delta\phi(\tau,\vec x)$, the equation 
of motion  reads 
\begin{align}
\phi_0''+\left(f_1 \frac{g'_0}{g_0} -2\frac{b'}{b}\right)\phi_0'+b^2 V'(\phi_0)=0 \label{KG0}
\end{align}
where here again $V'=dV/d\phi$  and to first order in perturbations we get
\begin{align}
\delta \phi''&-\frac{b^2}{a^2}\nabla^2 \delta\phi +b^2 V''(\phi_0)\delta\phi+\left(f_1 \frac{g'_0}{g_0} -2\frac{b'}{b}\right)\delta\phi'\nonumber \\
&+\left(
f_1f_2\frac{g'_0}{g_0}\frac{\delta g}{g_0}+f_1\left(\frac{\delta g}{g_0}\right)'-2\Phi'-\nabla^2B\right)\phi'_0\nonumber \\
&+2\Phi b^2V'(\phi_0)=0\label{KG1}
\end{align}
where we have used the notation
\begin{align}
f_1=\frac{d\ln f}{d\ln g}
\end{align}
and
\begin{align}
f_2=\frac{d\ln f_1}{d\ln g}
\end{align}
The perturbation of the determinant reads
\begin{align}
\frac{\delta g}{g_0}= 2(\nabla^2 E+\Phi-3\Psi)\label{gpert}
\end{align}
with $g_0=a^6b^2$.

On the other hand, the perturbed Einstein equations 
$\delta G^\mu_{\;\;\nu}=8\pi G\, \delta T^{\mu}_{\;\;\nu}$ read
\begin{align}
&\frac{2}{b^2}\left[\frac{a'}{a}(\nabla^2(E'-B)-3\Psi')-3\frac{a'^2}{a^2}\Phi+\frac{b^2}{a^2}\nabla^2\Psi\right]\nonumber \\
&=8 \pi G \delta T^{0}_{\;\; 0}\label{E00} \\
&\frac{2}{b^2}\left[\Psi'+\frac{a'}{a}\Phi\right]_{,i}=8 \pi G \delta T^{0}_{\;\; i}\label{E0i} \\
&\frac{2}{b^2}\left[-\Psi''+\frac{b'}{b}\Psi'-\frac{a'}{a}(\Phi'+3\Psi')-\frac{a'^2}{a^2}\Phi\right.\nonumber \\
&\left.+\frac{b^2}{2a^2}\nabla^2(\Psi-\Phi)+2\left(\frac{b'}{b}\frac{a'}{a}-\frac{a''}{a}\right)\Phi\right.\nonumber\\
&\left.+\frac{1}{2}\nabla^2(E''-B')-\frac{1}{2}\left(\frac{b'}{b}-3\frac{a'}{a}\right)\nabla^2(E'-B)\right]\delta^i_{\;j}
\nonumber\\
&-\frac{2}{b^2}D_{,ij}=8 \pi G \delta T^i_{\; j}
\end{align}
where
\begin{align}
D=\frac{b^2}{a^2}(\Phi-\Psi)+ \frac{1}{2} (E''-B')-\frac{1}{2}\left(\frac{b'}{b}-3\frac{a'}{a}\right)
(E'-B)
\end{align}
and the energy-momentum tensor components read
\begin{align}
\delta T^0_{\;0}&=\delta\rho=\frac{f}{\sqrt{g}}\left[2(1-f_1)\left(\frac{\phi_0'\delta\phi'}{b^2}+\frac{{\phi_0'}^2}{b^2}\left[-\Phi
\right.\right.\right.\nonumber \\
+&\left.\left.\left.\frac{1}{2}\left(f_1-\frac{1}{2}\right)\frac{\delta g}{g_0}\right]\right)+2f_1 \left(V'\delta\phi+ V\left(f_1-\frac{1}{2}\right)\frac{\delta g}{g_0}\right)\right.\nonumber \\
&\left.-f_1f_2\left(\frac{{\phi_0'}^2}{b^2}-2V\right)\frac{\delta g}{g_0}\right]\label{deltarho}\\
\delta T^0_{\;i}&= \frac{1}{b^2}\frac{f}{\sqrt{g}}\phi_0'\delta\phi_{,i}\\
\delta T^i_{\;j}&=-\delta p\;\delta^i_j=-\frac{f}{\sqrt{g}}\left[2f_1\left(\frac{\phi_0'\delta\phi'}{b^2}+\frac{{\phi_0'}^2}{b^2}\left[-\Phi\right.\right.\right.\nonumber \\
+&\left.\left.\left.\frac{1}{2}\left(f_1-\frac{1}{2}\right)\frac{\delta g}{g_0}\right]\right)-2f_1 \left(V'\delta\phi+ V\left(f_1-\frac{1}{2}\right)\frac{\delta g}{g_0}\right)\right.\nonumber \\
&+\left. f_1f_2\left(\frac{{\phi_0'}^2}{b^2}-2V\right)\frac{\delta g}{g_0}\right]\delta^i_j 
\end{align}

Notice that the $i\neq j$ equations imply
\begin{align}
D=0\label{D}
\end{align}
We can also write  
\begin{align}
\delta T^{0}_{\;\; i}=\frac{1}{b^2}\frac{f}{\sqrt{g}}\phi_0'\delta\phi_{,i}=(\rho + p)\left(\frac{\delta\phi}{\phi_0'}\right)_{,i}
\end{align}
In the rest gauge in which $\delta \phi=0$, we have $\delta T^{0}_{\;\; i}=0$ and we can write the pressure perturbation as 
\begin{align}
\delta p&=\frac{f_1}{1-f_1}\delta \rho  +\left(f_1-\frac{1}{2}\right)\left(w_\phi-\frac{f_1}{1-f_1}\right) \rho_0\frac{\delta g}{g_0}\nonumber \\
&+\frac{w_\phi f_2}{1-f_1}\rho_0\frac{\delta g}{g_0} \label{dp}
\end{align}
where $w_\phi$ is the background equation of state. Thus we see that unlike the Diff invariant case \cite{Gordon:2004ez}  in which in the rest frame $\delta p=c_s^2 \delta \rho$ and $c_s^2=1$, now we have additional  contributions from metric perturbations to the pressure perturbation. 

\subsection{Adiabatic perturbations in models with $V=0$}

Let us consider models with $V=0$ and constant $f_1=\alpha$ i.e. $f_2=0$. In such a case 
$\omega_\phi=\alpha/(1-\alpha)$ and from \eqref{dp} we have $\delta p=w_\phi \delta \rho$, i.e.  the speed of sound in such a case is simply $c_s^2=w_\phi$  and we would have  adiabatic pressure perturbations.

On the other hand, Bianchi identities imply  $T^\mu_{\;\;\nu ; \mu}=0$. In particular for $\nu=0$ we get at the background level
\begin{align}
(\alpha - 1)\phi_0''+ \left(\alpha - \frac{3}{2}\right)\left((\alpha - 1)\frac{b'}{b} + 3\alpha\frac{a'}{a}\right)\phi_0'=0 \label{backKG}
\end{align}
Imposing the field equation of motion \eqref{KG0}, we find
\begin{align}
\left(\alpha-\frac{1}{2}\right)\left((\alpha -1)\frac{b'}{b}+3\alpha\frac{a'}{a}\right)=0 \label{bp}
\end{align}
which is satisfied for the Diff invariant case $\alpha=1/2$ as expected, or for
\begin{align}
\frac{b'}{b}=-3\frac{\alpha}{(\alpha -1)}\frac{a'}{a} \label{aabb}
 \end{align} 
and therefore \eqref{backKG} implies for $\alpha \neq 1$
\begin{align}
\phi_0''=0\label{phi0}
\end{align}
as expected from \eqref{phiprima}. 
To first order in metric perturbations the $\nu=0$ and $\nu=i$ conservation equations imply, using \eqref{phi0} and \eqref{bp}
\begin{align}
&(\alpha-1)\delta \phi''+\frac{b^2}{2a^2}\nabla^2 \delta\phi
+\left(\left(\alpha-\frac{3}{2}\right)\left((\alpha-1) \Phi' -3\alpha\Psi'\right.\right.\nonumber \\
+&\left.\left.\alpha\nabla^2E'\right)+\frac{1}{2}\nabla^2B\right)\phi_0'=0\label{dp0}\\
&\delta\phi'+\left((\alpha-1) \Phi -3\alpha\Psi+\alpha\nabla^2E\right)\phi_0'=0\label{dpi}
\end{align}
It can be seen that using the second equation on the first one we recover the perturbed field equation \eqref{KG1} 

On the other hand, using the gauge condition $E'-B=0$ in \eqref{D}, we get
\begin{align}
\Phi=\Psi
\end{align}
Notice that these results are also valid when in addition to the TDiff scalar field  we also have a  Diff invariant matter sector since the condition imposed by the conservation of the total energy-momentum tensor only affects the TDiff part as  the energy-momentum tensor of the Diff invariant sector  is automatically conserved.

Now combining the first and third Einstein equations we get
\begin{align}
(\alpha - 1)\Phi'' +\alpha \frac{b^2}{a^2}\nabla^2\Phi+4\Phi' \frac{a'}{a}(\alpha - 1)=0 \label{Phie}
\end{align}
In the super-Hubble regime $k\ll aH$ the second term can be neglected and we obtain for $\alpha\neq 1$
the simple expression
\begin{align}
\Phi_k'' +4\frac{a'}{a}\Phi_k'=0 
\end{align}
which is independent of $\alpha$.

Changing to conformal time $a(\eta)d\eta=b(\tau)d\tau$ in \eqref{Phie} and using \eqref{aabb} we get in Fourier space
\begin{align}
\Phi_k'' + 3(1+c_s^2)\frac{a'}{a}\Phi'_k +c_s^2 k^2 \Phi_k=0 \label{Phik}
\end{align}
where $'$ now denotes derivative with respect to the conformal time.  Notice that this is the same equation obtained for adiabatic perturbations  in a Diff invariant theory \cite{Mukhanov}. 

This in particular implies that for positive energy density i.e. $\alpha<1$, we have a constant and a decreasing mode so that at long times
in the super-Hubble regime, i.e.,  $\Phi_k=C_k$ with $C_k$ constant. 
In particular, the curvature perturbation defined as 
\begin{align}
\zeta_k=\frac{2}{3(1+w_\phi)}\left(\frac{\Phi'_k}{{\cal H}}+\Phi_k\right)+\Phi_k
\end{align}
with ${\cal H}=a'/a$ is conserved on super-Hubble scales, i.e
\begin{align}
\zeta'_k=0 \;\; (k\eta\ll 1)
\end{align}

Also in the super-Hubble regime, we get for the constant mode
using \eqref{E00}, \eqref{dpi} and \eqref{E0i}
\begin{align}
\nabla^2E=4\Phi_k
\end{align}
and
\begin{align}
\frac{\delta\phi'_k}{\phi'_0}=(1-2\alpha)\Phi_k
\end{align}
so that as expected
\begin{align}
\frac{\delta\rho_k}{\rho_0}=-2\Phi_k
\end{align}
which agrees with the result for  perfect fluids  in the longitudinal gauge  in a Diff invariant theory \cite{Mukhanov}.

\section{TDiff invariant dark matter}
Let us consider the simple model
\begin{align}
S_{\phi}= \int d^4x\; \frac{1}{2}g^{\mu \nu}\partial_\mu\phi\,\partial_\nu\phi
\end{align}
which corresponds to $f(g)=1$, i.e. $\alpha=0$. The 
equation of state would be  $w_\phi=c_s^2=0$ which corresponds to non-relativistic matter. 
In this case, we have from \eqref{aabb} $b=$const.
Considering the matter dominated era,  equation \eqref{Phik} reads in 
conformal time
\begin{align}
\Phi_k'' + 3\left(\frac{a'}{a}\right)\Phi'_k=0
\end{align}
whose general solution reads
\begin{align}
\Phi_k(\eta)=A_k+\frac{C_k}{\eta^5}
\end{align}
so that at late times the constant mode dominates.  From \eqref{dpi} with 
$\alpha=0$ we get 
\begin{align}
\frac{\delta\phi'_k}{\phi'_0}=\Phi_k
\end{align}
which implies $\delta\phi'_k$ is constant at late times. 
On the other hand, from \eqref{dp0} we get in conformal time 
\begin{align}
\frac{1}{a}(a\nabla^2 E_k')'=k^2\frac{\delta\phi'_k}{\phi'_0}=k^2\Phi_k
\end{align}
so that integrating in time, the dominant contribution  at late
times reads
\begin{align}
\nabla^2 E_k=\frac{k^2\eta^2}{6}\Phi_k
\end{align}
so that from \eqref{deltarho} for the density contrast in the sub-Hubble regime $k\eta \gg 1$  we recover the standard expression for a 
matter fluid in a Diff invariant theory
\begin{align}
\frac{\delta\rho_k}{\rho_0}=-\nabla^2 E_k=-\frac{k^2\eta^2}{6}\Phi_k \propto a
\end{align}
which grows with the scale factor as expected for a non-relativistic matter component.  

Notice that in this case with massless scalar fields which are the only source of the gravitational fields, the time scale for the gravitational field evolution is similar to that of the scalar field, i.e. we are not in the geometric optics regime described in Section III. Accordingly, the results for the propagation speed of scalar field disturbances equal to the speed of light obtained there do not apply in this context. 

\section{Conclusions}
In this work we have explored the effects of the breaking of Diff invariance in the matter sector in cosmological contexts. We have considered simple scalar field models which break Diff invariance down to the TDiff subgroup. We have analyzed the viability of those models and shown that the potential issues related to violations of the weak equivalence principle and the difference between inertial and passive gravitational masses can be avoided in models involving a global modification of the integration measure
in the action. In those cases, TDiff models behave in the same way as 
the corresponding Diff models in the geometric optics approximation, the only difference being the relation between the amplitude of the field and the number density of particles which is rescaled by a $f/\sqrt{g}$ factor.  However, we have shown that beyond the geometric optics approximation, for example, for homogeneous fields  in cosmological contexts, the scaling of the energy density can be drastically different.  This implies that the typical stiff fluid behaviour of kinetically dominated homogeneous scalar fields is no longer satisfied and a rich phenomenology opens up  in this class of models. Thus we have presented examples of inflationary models driven by canonical kinetic terms, which are able to generate a nearly scale-invariant (red-tilted) spectra of density fluctuations or dark matter models described by massless scalar fields with the correct clustering properties. Slow-roll parameters are also modified in this class of models allowing in principle for any potential term to support inflation provided the $f/\sqrt{g}$ factor is large enough in the early universe.

The analysis of cosmological perturbations shows that Diff invariance breaking generically introduces a new  source of non-adiabaticity, 
although the simplest power-law models without potential terms 
describe adiabatic perfect fluids.

The new phenomenology related to these models could make them 
useful tools for the description of the dark or inflationary sectors of cosmology.
Moreover, the fact that the observable modifications only affects very-low frequency modes justifies the analysis of TDiff modifications of the Standard Model of elementary particles.  Work is in progress in this direction.


\acknowledgements{This paper is dedicated to the memory of Francisco L\'opez Secilla. I would like to thank the anonymous referee  for suggesting the introduction of the covariant formalism. I also thank David Alonso L\'opez, Gerardo Garc\'{\i}a-Moreno  and Javier de Cruz P\'erez for useful comments and suggestions. I would also like to thank Antonio Miguel Gonz\'alez Bello-Morales, Mar\'{\i}a del Prado Mart\'{\i}n Moruno, Dar\'{\i}o Jaramillo and Alfredo Delgado Miravet for useful discussions about TDiff theories.   This work has been supported by the MICIN (Spain) projects PID2019-107394GB-I00 (AEI/FEDER, UE) and PID2022-138263NB-I00 (AEI/FEDER, UE).}

\section{Appendix: transverse diffeomorphisms}
In this Appendix we will review the fundamentals of  transverse diffeomorphisms and introduce some of the notation that we  use in the rest of the work. 

Let us consider an infinitesimal general coordinate transformation given by
\begin{align}
\hat x^\mu=x^\mu+\xi^\mu(x)\label{cotr}
\end{align}
The corresponding metric tensor transformation will read
\begin{align}
\hat g_{\mu\nu}(\hat x)=\frac{\partial x^\alpha}{\partial \hat x^\mu}\frac{\partial x^\beta}{\partial \hat x^\nu}g_{\alpha\beta}(x)
\end{align}
so that the metric determinant  $g(x)=\vert \det g_{\mu\nu}(x)\vert$ transforms 
as
\begin{align}
\hat  g(\hat x)=\left\vert \frac{\partial x^\alpha}{\partial \hat x^\mu}\right\vert^2 g(x) = g(x)(1-2\partial_\mu \xi^\mu(x))
\end{align}
Coordinate transformations satisfying the condition
\begin{align}
\partial_\mu \xi^\mu=0
\end{align}
thus preserve the metric determinant. Notice that they  also preserve the volume element $d^4 \hat x= d^4 x$ and accordingly any action term of the form 
\begin{align}
S=\int d^4x f(g){\cal L}(x)\label{TDiffaction}
\end{align}
where $f(g)$ is an arbitrary function of the metric determinant and ${\cal L}(x)={\cal L}(\phi(x),\partial_\mu\phi(x),g_{\mu\nu}(x))$ is a scalar function of the matter fields $\phi(x)$, its derivatives and the metric, is invariant under volume preserving coordinate transformations, i.e.
\begin{align}
\hat S=\int d^4\hat x f(\hat g(\hat x)){\cal L}(\hat x)=\int d^4 x f(g(x)){\cal L}(x)=S
\end{align}

The same conclusions can be obtained from the point of view of gauge transformations (diffeomorphisms). Let us consider the gauge transformations induced on the metric tensor by the  coordinate transformations \eqref{cotr} 
 \begin{align}
 \delta g_{\mu\nu}(x)=\hat g_{\mu\nu}(x)-g_{\mu\nu}(x)=-\xi_{\mu ; \nu}-\xi_{\nu ; \mu}
 \end{align}
The metric determinant will change accordingly as
 \begin{align}
  \delta g(x)=\hat g(x)-g(x)=-2g(x)\xi^\mu_{\; ;\mu}
   \end{align}
The Lagrangian density ${\cal L}(x)$ being a scalar field transforms as
\begin{align}
\delta{\cal L}(x)&=\hat{\cal L}(x)-{\cal L}(x)=-\xi^\mu(x) \partial_\mu {\cal L}(x)
\end{align}
On the other hand
$f(g)$ transforms as
\begin{align}
\delta f(g(x))&=f'(g(x))\delta g(x)=-f'(g(x))2g(x)\xi^\mu_{\; ;\mu}
\end{align}
where prime denotes derivative with respect to its argument. 
Thus the variation of the action reads
\begin{align}
\delta S&=\int d^4 x \left(-f'(g)2g\, \xi^\mu_{\; ;\mu} {\cal L}-f(g)\xi^\mu\partial_\mu {\cal L}\right)
\nonumber\\
&=\int d^4x  \left(-2f'(g)\sqrt{g}\partial_\mu (\sqrt{g}\xi^\mu) {\cal L}-f(g)\xi^\mu\partial_\mu {\cal L}\right)
\end{align}
Integrating by parts in the second term we find
\begin{align}
\delta S&=\int d^4x  \left(-f'(g)\partial_\mu g \xi^\mu {\cal L}- 2gf'(g)\partial_\mu\xi^\mu {\cal L}\right.\nonumber \\
&\left.+f'(g)\partial_\mu g \xi^\mu {\cal L}+f(g)\partial_\mu\xi^\mu {\cal L}\right)\nonumber \\
&= 
\int d^4x \; \partial_\mu\xi^\mu ( f(g)-2gf'(g)) {\cal L}
\end{align}
Thus we see that the action is invariant for arbitrary $\xi^\mu$ only in the case in which 
$f(g)-2gf'(g)=0$ which corresponds to $f\propto \sqrt{g}$ as expected. However it is also invariant for arbitrary $f(g)$ for transverse diffeomorphisms satisfying $\partial_\mu\xi^\mu=0$.

\end{document}